\documentclass[structabstract]{aa}  
\usepackage{graphicx}
\usepackage{txfonts}
\usepackage[round]{natbib}
\bibpunct{(}{)}{;}{a}{}{,} 
\usepackage{epsfig}

\begin{document}
\title{The outburst and nature of two young eruptive stars\\ in the
  North America/Pelican Nebula Complex}

   \author{\'A. K\'osp\'al\inst{1}
          \and
          P. \'Abrah\'am\inst{2}
          \and
          J. A. Acosta-Pulido\inst{3,4}
          \and
          M. J. Ar\'evalo Morales\inst{3,4}
          \and
          M. I. Carnerero\inst{3,4}
          \and
          E. Elek\inst{2}
          \and
          J. Kelemen\inst{2}
          \and
          M. Kun\inst{2}
          \and
          A. P\'al\inst{2,5}
          \and
          R. Szak\'ats\inst{6}
          \and
          K. Vida\inst{2}
          }

   \institute{Leiden Observatory, Leiden University,
              PO Box 9513, 2300 RA Leiden, The Netherlands\\
              \email{kospal@strw.leidenuniv.nl}
          \and
              Konkoly Observatory of the Hungarian Academy of Sciences,
              PO Box 67, 1525 Budapest, Hungary
          \and
              Instituto de Astrof\'\i{}sica de Canarias, 38200 La
              Laguna, Tenerife, Spain
          \and
              Departamento de Astrof\'\i{}sica, Universidad de La Laguna,
              38205 La Laguna, Tenerife, Spain
          \and
              Department of Astronomy, Lor\'and E\"otv\"os University,
              P\'azm\'any P. st. 1/A, Budapest H-1117, Hungary
          \and
              Baja Astronomical Observatory of B\'acs-Kiskun County,
              PO Box 766, 6500 Baja, Hungary 
              }

   \date{Received date; accepted date}

 
   \abstract{In August 2010, the sudden optical brightening of two
     young stellar objects, HBC\,722 and VSX\,J205126.1+440523,
     located in the North America/Pelican Nebula Complex, was
     announced. Early photometric and spectroscopic observations of
     these objects indicated that they may belong to the FUor or EXor
     class of young eruptive stars. The eruptions of FUors and EXors
     are often explained by enhanced accretion of material from the
     circumstellar disk to the protostar.}
   {In order to determine the true nature of these two objects, we
     started an optical and near-infrared monitoring program, and
     complemented our data with archival observations and data from
     the literature.}
   {We plot and analyze pre-outburst and outburst spectral energy
     distributions (SEDs), multi-filter light curves, and color-color
     diagrams.}
   {The quiescent SED of HBC\,722 is consistent with that of a
     slightly reddened normal T\,Tauri-type star. The source
     brightened monotonically in about two months, and the SED
     obtained during maximum brightness indicates the appearance of a
     hot, single-temperature blackbody. The current fading rate
     implies that the star will return to quiescence in about a year,
     questioning its classification as a bone fide FUor. The quiescent
     SED of VSX\,J205126.1+440523 looks like that of a highly embedded
     Class I source. The outburst of this source happened more
     gradually, but reached an unprecedentedly high amplitude. At 2.5
     months after the peak, its light curves show a deep minimum, when
     the object was close to its pre-outburst optical
     brightness. Further monitoring indicates that it is still far
     from being quiescent.}
   {The shape of the light curves, as well as the bolometric
     luminosities and accretion rates suggest that these objects do
     not fit into the classic FUor group. Although HBC\,722 exhibit
     all spectral characteristics of a bona fide FUor, its luminosity
     and accretion rate is too low, and its timescale is too fast
     compared to classical FUors. VSX\,J205126.1+440523 seems to be an
     example where quick extinction changes modulate the light curve.}


   \maketitle


\section{Introduction}

In August 2010, two new young eruptive star candidates were discovered
in the North America/Pelican Nebula Complex (distance: 550\,pc,
\citealt{straizys1989}). HBC\,722 (also known as LkH$\alpha$\,188\,G4
and PTF\,10qpf) brightened by $\Delta$R=3.3\,mag between 2010 May and
August \citep{semkov2010a}. VSX\,J205126.1+440523 (also known as
IRAS\,20496+4354 and PTF\,10nvg) brightened by 1.8\,mag in unfiltered
light between 2009 December and 2010 June, but Digitized Sky Survey
plates show that it had been several magnitudes fainter in quiescence
\citep{itagaki2010,munari2010}. \citet{semkov2010} and
\citet{miller2010} provided light curves and spectroscopy for HBC\,722
and concluded that we witness a bona fide outburst of a FUor-type
object. FUors, named after the prototype object FU\,Orionis, brighten
by up to 5\,mag at optical wavelengths and may stay in the high state
for decades. \citet{covey2010} presented light curves and spectroscopy
for VSX\,J205126.1+440523, and found that in many respects this object
is different from FUors or EXors (the latter being another class of
eruptive YSOs, named after the prototype EX\,Lup, which flares up by
1-5\,mag in every few years and stay bright for several
months). Currently only about two dozens of young eruptive stars
(FUors and EXors) are known, thus the two new outbursts announced in
August 2010 are noteworthy events. Should they turn out to be
accretion-powered eruptions, their detailed study may contribute to
the understanding of these important phases of early stellar
evolution. 

In this paper we present an optical and infrared view of the two
eruptive star candidates. Using archival and new data, we characterize
their circumstellar environment and compare them with those of some
better studied FUors and EXors. We present new optical and
near-infrared photometric data points taken during the outburst, which
indicate that HBC\,722 already passed its peak brightness and started
a monotonous fading with a steady fading rate, while neither the
brightening, nor the fading of VSX\,J205126.1+440523 was monotonous.


\section{Observations and data reduction}

\paragraph{Optical observations.}
(B)VRI-band images were obtained between 19 September 2010 and 2
January 2011 with three telescopes: the 60/90/180\,cm (aperture
diameter/primary mirror diameter/focal length) Schmidt telescope of
the Konkoly Observatory (Hungary), the 1\,m (primary mirror diameter)
RCC telescope of the Konkoly Observatory, and the 80\,cm (primary
mirror diameter) IAC-80 telescope of the Teide Observatory in the
Canary Islands (Spain). The Konkoly Schmidt telescope is equipped with
a 4096\,$\times$\,4096 pixel Apogee Alta U16 CCD camera (pixel scale:
1.03$''$), and a Bessel BV(RI)$_{\rm C}$ filter set. The 1\,m RCC is
equipped with a 1300\,$\times$\,1340 pixel Roper Scientific
WersArray:1300B CCD camera (pixel scale: 0.306$''$), and a Bessel
UBV(RI)$_{\rm C}$ filter set. The Teide IAC-80 telescope is equipped
with a 2048\,$\times$\,2048 pixel Spectral Instruments E2V 42-40
back-illuminated CCD camera `CAMELOT' (pixel scale: 0.304$''$), and a
Johnson-Bessel UBV(RI)$_{\rm J}$ filter set. The images were reduced
in IDL following the standard processing steps of bias subtraction and
flat-fielding. On each night, for each target, images were obtained in
blocks of 3 or 5 frames per filter. Aperture photometry for the target
and other field stars were performed on each image using IDL's
\textit{cntrd} and \textit{aper} procedures. Since HBC\,722 is
surrounded by a reflection nebula, in order to be consistent with the
photometry of \citet{semkov2010}, we used the same apertures: an
aperture radius of 4$''$ and a sky annulus between 13$''$ and
19$''$. For HBC\,722, instrumental magnitudes were transformed to the
standard system using the 8 brightest stars (from star `A' to star
`H') from the comparison sequence given in \citet{semkov2010}. For
each image we fitted the difference of the instrumental and the
standard magnitudes of the comparison stars as a function of the V$-$I
color, and used this relationship to convert the instrumental
magnitudes of our target to standard magnitudes. For
VSX\,J205126.1+440523, we observed the standard field NGC\,6823 with
the Schmidt telescope during the photometric night 23/24 September
2010, and calibrated 6 comparison stars in the vicinity of our
target. A finding chart and the standard magnitudes of the comparison
stars can be seen in Fig.~\ref{fig:compstars} and in
Tab.~\ref{tab:compstars} in the Online Material. The conversion of
instrumental to standard magnitudes was done the same way as for
HBC\,722. Similarly to the comparison stars of HBC\,722, we cannot
exclude that the comparison stars of VSX\,J205126.1+440523 might be
variables on longer timescales, although they were constant within the
measurement uncertainties during our observing period. The resulting
photometry for our two targets is presented in
Tabs.~\ref{tab:phot_vsxj20} and \ref{tab:phot_hbc722} in the Online
Material. We note that the R and I filters on the two telescopes are
different, which may introduce a systematic difference in the
magnitudes obtained with the different telescopes. However, in our
experience, this difference is less than 0.05\,mag
\citep{kospal2010}. Since the observed brightness variations of our
targets are much larger than 0.05\,mag, this possible difference in
the filter systems does not affect our analysis and conclusions.

\paragraph{Near-infrared observations.} JHK$_{\rm S}$ images were
obtained with the 1.52\,m Telescopio Carlos Sanchez (TCS, Teide
Observatory, Spain) between 19 September and 19 November 2010, using
the 256$\,{\times}\,$256 pixel Nicmos 3 detector CAIN-3 with the wide
field optics (pixel scale: 1$''$). Observations were performed in a
5-point dither pattern in order to enable proper sky subtraction. The
images were reduced using a modified version of {\it caindr}, an iraf
data reduction package written by J.~Acosta-Pulido\footnote{for more
  details on the {\it caindr} package, see
  http://www.iac.es/telescopes/cain/cain\_eng.html.}. Data reduction
steps included sky subtraction, flat-fielding, and the co-addition of
all frames taken with the same filter. The sky image was obtained as
the median combination of all frames, masking regions occupied by
bright sources. The final image was produced using the standard
``shift-and-add'' technique, including rejection of outlier
pixels. The instrumental magnitudes of the target and all good-quality
2MASS stars in the field were extracted using aperture photometry in
IDL. For the photometric calibration we used the Two Micron All Sky
Survey (2MASS) catalog \citep{cutri2003}. We determined the offset
between the instrumental and the 2MASS magnitudes by averaging
typically 20-30 stars, using {\it biweight\_mean}, an
outlier-resistant averaging method. The resulting photometry of our
two targets is presented in Tabs.~\ref{tab:phot_vsxj20} and
\ref{tab:phot_hbc722} in the Online Material. We obtained additional
near-infrared photometry using archival images from the UKIRT InfraRed
Deep Sky Surveys (UKIDSS). These images were taken with the Wide Field
Camera on the 3.8\,m diameter UKIRT in 2006, and they are part of Data
Release 8. Aperture photometry and calibration were executed in the
same way as for the TCS data.


\section{Results and analysis}


\subsection{HBC\,722}

HBC\,722 is part of a small cluster of young stars called
``LkH$\alpha$\,188 group'' by \citet{cohen1979}. In their naming
convention, HBC\,722 is called LkH$\alpha$ 188 G4. The whole cluster
is located in a dark cloud separating the North America and the
Pelican Nebulae \citep{straizys1989}. In outburst, the star is
surrounded by a compact, asymmetric reflection nebula, which is well
visible at optical wavelengths \citep{miller2010}, but also
discernible in our J band images.

\begin{figure}
\centering \includegraphics[width=\columnwidth]{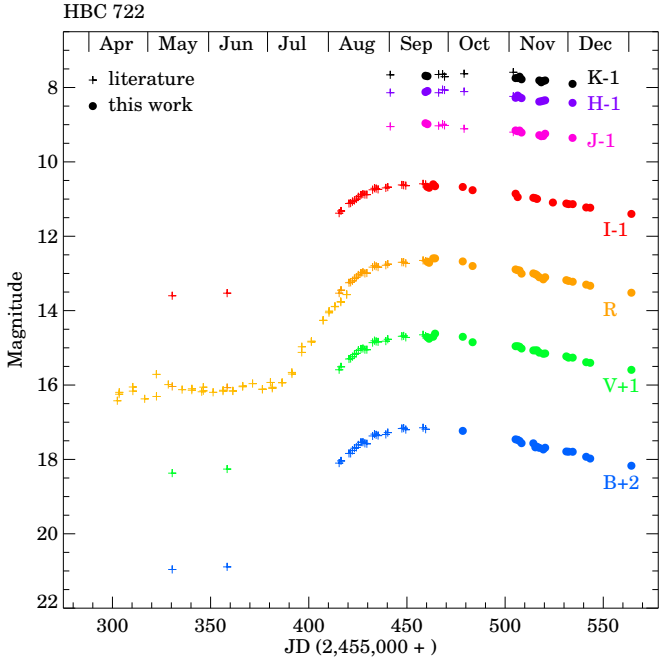}
\caption{Light curves of HBC\,722. For better visibility, the B, V, I,
  J, H, and K$_{\rm S}$ light curves were shifted along the y axis by
  the values indicated in the figure. Filled dots are from this work,
  plus signs are from \citet{semkov2010} and \citet{miller2010}.}
\label{fig:light_hbc722}
\end{figure}

\paragraph{Light curve.} The brightening of HBC\,722 is well
documented in \citet{semkov2010} and \citet{miller2010}. Between
August 2009 and July 2010, the star gradually brightened by 1\,mag in
the R band, then between July and August 2010, it brightened by
another 3\,mag, reaching a maximum brightness at the end of August
2010. We have been monitoring the star since September 2010. Our data
confirm that the star reached is maximal brightness, and is currently
gradually fading (Fig.~\ref{fig:light_hbc722}). Between 20 September
2010 and 3 December 2010, HBC\,722 decreased its brightness by 0.55,
0.54, 0.47, 0.37, 0.32, and 0.21\,mag in the B, V, R, I, J, H, and
K$_{\rm S}$ bands, respectively, indicating that the source has become
slightly redder. Fitting a line to the data points between these two
epochs gives fading rates of 0.34, 0.25, 0.25, 0.21, 0.16, 0.13, and
0.07 mag/month in the B, V, R, I, J, H, and K$_{\rm S}$ band,
respectively. Our last optical data points taken on 2 January 2011 fit
into this trend. Assuming that the linear fading continues with these
rates, and taking into account the pre-outburst optical fluxes
observed by \citet{semkov2010} around July 2009, we estimate that the
source will return to quiescence some time between fall 2011 and
spring 2012. The two pre-outburst JHK$_{\rm S}$ data sets (2MASS from
June 2000 and UKIDSS from July 2006) agree within 0.1\,mag, indicating
a rather constant pre-outburst near-infrared brightness. Considering
the relatively slow near-infrared fading rates, HBC\,722 may exhibit
higher than quiescent near-infrared fluxes even until fall 2013. Thus,
current fading rates indicate that the outburst of HBC\,722 would last
approximately 2 or 3 years.

\begin{figure}
\centering \includegraphics[width=\columnwidth]{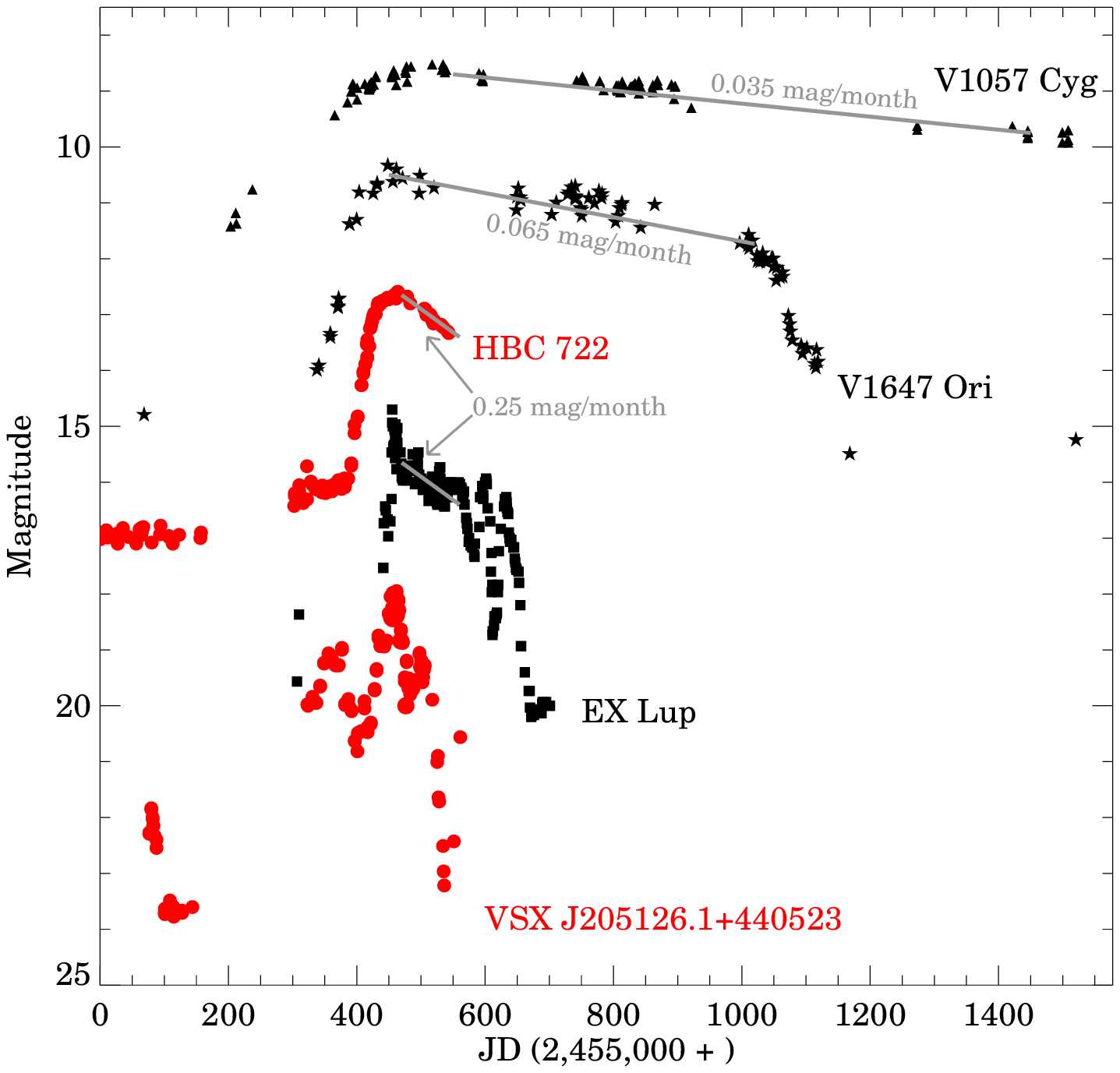}
\caption{Light curves of our targets and those of different young
  eruptive stars. {\it Triangles:} photographic light curve of
  V1057\,Cyg from \citet{gieseking1973}; {\it stars:} I$_{\rm C}$
  light curve of V1647\,Ori from \citet{acosta2007}; {\it squares:}
  visual light curve of EX\,Lup during its extreme outburst in 2008
  \citep{abraham2009}. For HBC\,722 and VSX\,J205126.1+440523 we
  plotted the R-band light curves. The data for V1057\,Cyg,
  V1647\,Ori, and EX\,Lup were shifted along the y axis for better
  visibility and also along the x axis so that the peak brightness is
  approximately at the same position for all stars.}
\label{fig:context}
\end{figure}

In order to put into context the outburst history of HBC\,722, in
Fig.~\ref{fig:context} we compare its light curve with those of other
young eruptive objects. The recent extreme outburst of EX\,Lup
exhibited a faster onset, more peaked maximum, and a fading which was
initially very similar to that of HBC\,722. However, EX\,Lup showed
two deep minima before it went back to quiescence some 8 months after
its peak brightness. The light curve of V1057\,Cyg, the FUor having
the fastest known outburst and fading, is still much slower than that
of HBC\,722. The light curve of V1647\,Ori, a source often classified
as an intermediate-type between FUors and EXors, initially displayed a
relatively slow fading, then suddenly went back to quiescence, thus
had an approximately 2.5-year-long outburst. If HBC\,722 continues the
linear fading it currently displays, the predicted outburst length
will be remarkably similar to that of V1647\,Ori. This comparison
implies that HBC\,722 is different from the classical FUors and is
more similar to EXors, or intermediate objects between FUors and
EXors.

\begin{figure}
\centering \includegraphics[width=\columnwidth]{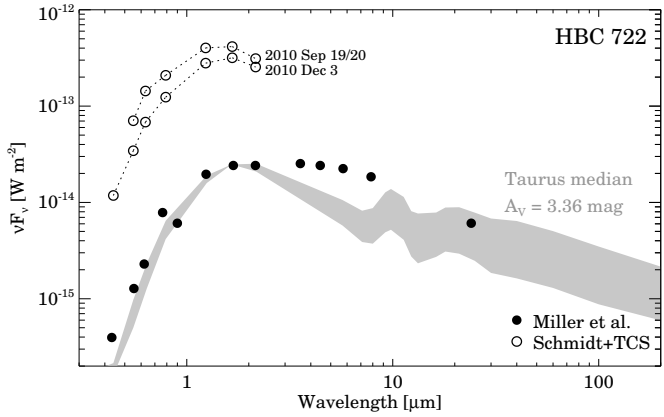}
\caption{Spectral energy distribution of HBC\,722. Filled dots are
  pre-outburst data from \citet{miller2010}, while open
  circles are outburst data (this work). The gray shaded area
  indicates the median SED of T\,Tauri stars in the Taurus star-forming
  region with spectral types between K5 and M2 (data below
  1.25$\,\mu$m and above 40$\,\mu$m are from \citealt{dalessio1999},
  data between 1.25 and 40$\,\mu$m are from \citealt{furlan2006}).}
\label{fig:sed_hbc722}
\end{figure}

\paragraph{Spectral energy distribution.} Fig.~\ref{fig:sed_hbc722}
shows the pre-outburst and outburst SEDs of HBC\,722. Pre-outburst
data are from \citet{miller2010} and references therein, while the
outburst data are from this work. In this figure, we also overplotted
with gray shading the typical SED of a T\,Tauri-type star in the
Taurus star-forming region \citep{dalessio1999,furlan2006}, scaled to
the H-band data point, and reddened by A$_{\rm V}$=3.36\,mag
\citep{cohen1979}. The outburst photometry indicates that a hot
continuum is added to the quiescent SED. The B, V, R, I, J, H, and
K$_{\rm S}$ points indicate a blackbody-like spectrum at all epochs
during the outburst. We could fit these points with a single
temperature blackbody and obtained a temperature of 4000\,K (using
A$_{\rm V}$=3.36\,mag). With the assumption that the SED is similar to
the Taurus median above 24$\,\mu$m, we calculated a pre-outburst
bolometric luminosity of 0.85\,L$_{\odot}$ by integrating the
de-reddened SED between 0.44 and 200$\,\mu$m. The outburst bolometric
luminosity can be similarly calculated, but due to the lack of
mid-infrared data points, we can either assume a blackbody shape until
10$\,\mu$m and assume that the SED did not change above 10$\,\mu$m, or
assume that the SED changed self-similarly in the whole
2$-$200$\,\mu$m range. The former approach give L$_{\rm
  bol}$=8.7\,L$_{\odot}$, the latter, L$_{\rm
  bol}$=12\,L$_{\odot}$. The true outburst luminosity is probably
between these two values.

\subsection{VSX\,J205126.1+440523}

VSX\,J205126.1+440523 is situated in an isolated molecular cloud
located about 15$'$ southeast of the Pelican Nebula molecular cloud
\citep{bally2003}. The eastern rim of this small cloud is well visible
in the [SII] and H$\alpha$ images of \citet{bally2003}. This
morphology suggests that VSX\,J205126.1+440523 sits on the tip of a
column of dense material, out of which it had been born. Apart from
the H$\alpha$ emission from the rim, no extended emission seems to be
associated with the source, not even in outburst. \citet{bally2003}
discovered several Herbig-Haro objects in this area, and claim that
one of them, HH\,569, is possibly driven by VSX\,J205126.1+440523.

\begin{figure}
\centering \includegraphics[width=\columnwidth]{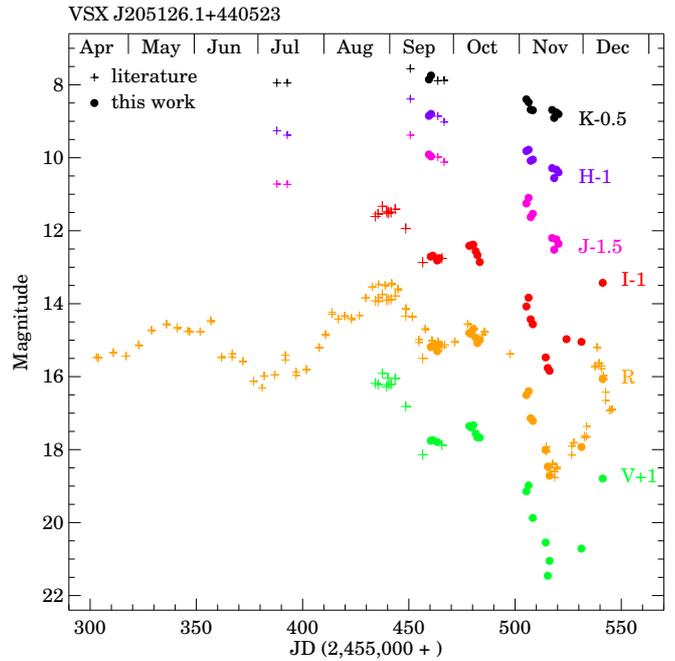}
\caption{Light curves of VSX\,J205126.1+440523. Filled dots are from
  this work, plus signs are from \citet{covey2010} and from Seiichiro
  Kiyota and Hiroyuki Maehara (vsnet,
  http://tech.dir.groups.yahoo.com/group/vsnet-recent-fuori/messages),
  crosses are visual estimates by \citet{itagaki2010}.}
\label{fig:light_vsxj20}
\end{figure}

\paragraph{Light curve.} In Fig.~\ref{fig:light_vsxj20} we plotted the
light curves of VSX\,J205126.1+440523. \citet{covey2010} reported the
source to be between R=18--19.25\,mag in mid-2009. After that, it
brightened by $\approx$6\,mag, reaching its maximal brightness in
August 2010. Since then, the source started fading, and by November
2010, it has nearly reached its mid-2009 optical brightness, then it
brightened again. The light curves show that neither the brightening,
nor the fading was monotonous. Although the near-infrared light curves
are not as well-sampled as the optical ones, they delineate similar
trends but with smaller amplitudes. We note that the R=19.25\,mag
reported by \citet{covey2010} may not be the true quiescent brightness
of the source, since the source was $\approx$20\,mag in the POSS2 red
plate taken in 1990 \citep{itagaki2010}. The comparison of the UKIDSS
and TCS photometry indicates that the source brightened by
$\Delta$J=7.9\,mag, $\Delta$H=6.7\,mag, and $\Delta$K$_{\rm
  S}$=4.8\,mag between July 2006 and September 2010. We note that the
source was K$_{\rm S}$=13.15\,mag in 2006, but it was not visible in
the K$_{\rm S}$ band in the 2MASS images taken in October 2000. Since
the 2MASS PSC is complete down to K$_{\rm S}$=14.3 \citep{cutri2003},
the source must have brightened at least 1.15\,mag between 2000 and
2006, making the true K$_{\rm S}$-band magnitude change at least
5.8\,mag.

The comparison of the light curve of VSX\,J205126.1+440523 with other
young eruptive stars in Fig.~\ref{fig:context} indicates that this
source is different from all the other sources plotted, although the
brightening and fading rates are most similar to those of
EX\,Lup. Especially remarkable is the deep minimum of
VSX\,J205126.1+440523 in November 2010, which is similar to the minima
displayed by EX\,Lup shortly before the end of the eruption.

\begin{figure}
\centering \includegraphics[width=\columnwidth]{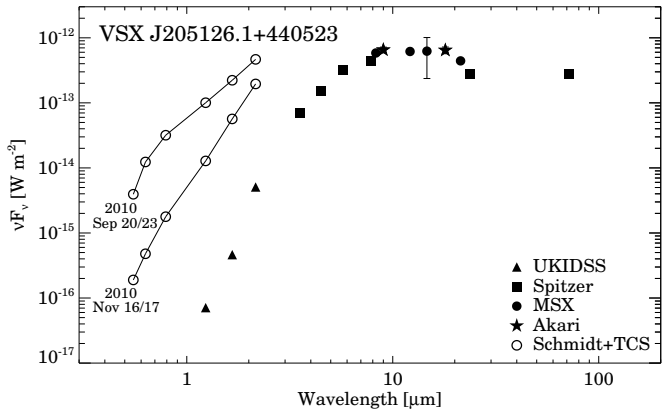}
\caption{Spectral energy distribution of VSX\,J205126.1+440523.
  Filled dots are pre-outburst data from various dates, while open
  circles are outburst data (see text).}
\label{fig:sed_vsxj20}
\end{figure}

\begin{figure}
\centering
\includegraphics[width=\columnwidth]{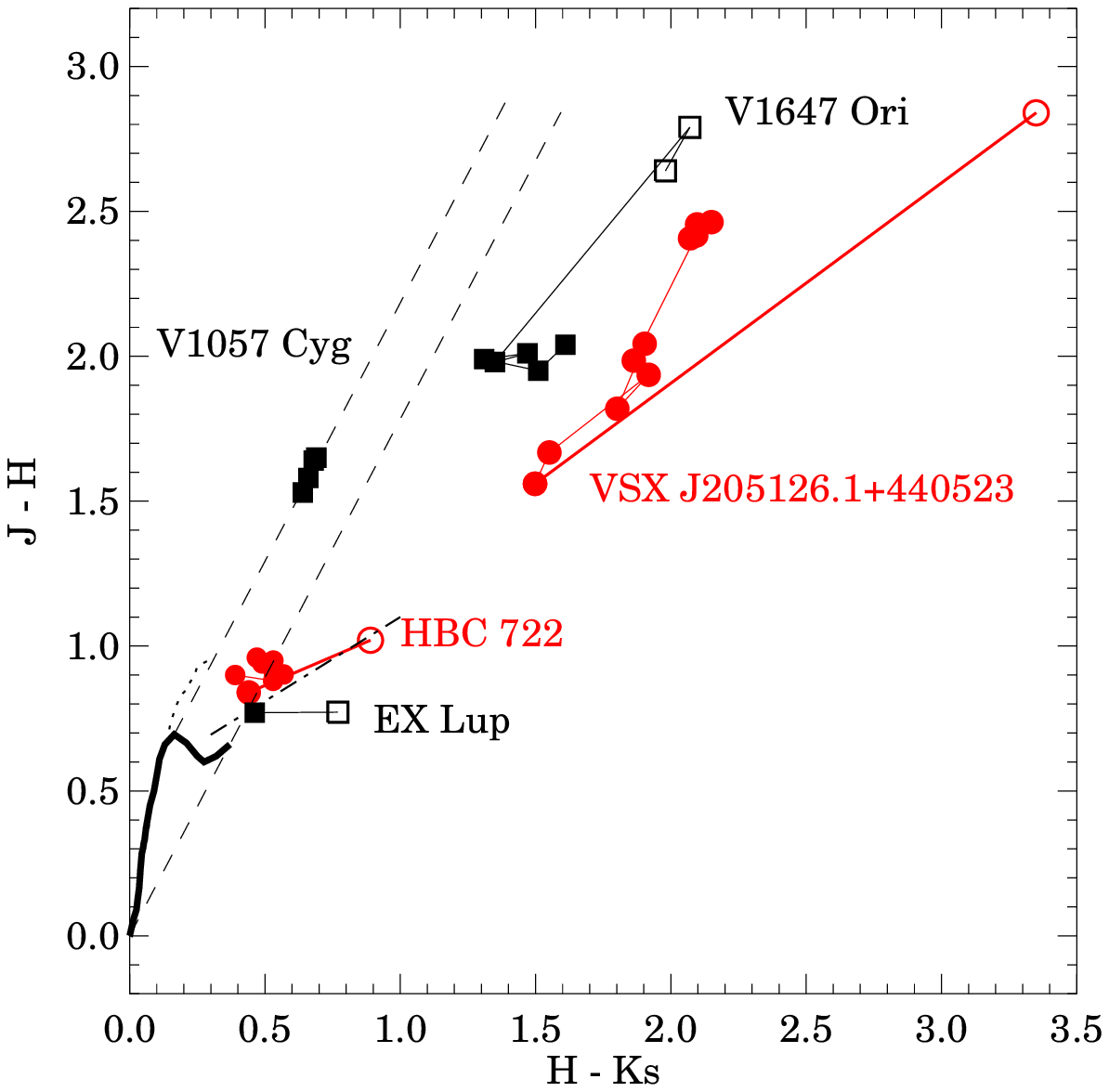}
\caption{Near-infrared color-color diagram. The main-sequence is
  marked by a thick solid line, the giant branch with dotted line
  \citep{koornneef1983}, the reddening path with dashed lines
  \citep{cardelli1989}, and the T\,Tauri locus with dash-dotted line
  \citep{meyer1997}. {\it Open symbols:} quiescent colors; {\it filled
    symbols:} outburst colors. Source of data: \citet{kenyon1991} for
  V1057\,Cyg, \citet{acosta2007} for V1647\,Ori, 2MASS PSC and
  \citet{juhasz2010} for EX\,Lup, and this work for HBC\,722 and
  VSXJ\,205126.1+440523.}
\label{fig:tcd}
\end{figure}

\paragraph{Spectral energy distribution.} In Fig.~\ref{fig:sed_vsxj20}
we compiled a pre-outburst SED using data from the UKIDSS database
(this work), the MSX6C Infrared Point Source Catalog \citep{egan2003},
the AKARI/IRC mid-infrared all-sky survey \citep{ishihara2010}, and
Spitzer data \citep[][and references therein]{covey2010}. Out of these
data points, the UKIDSS and the Spitzer are quasi-simultaneous (all
obtained between June and August 2006), while the MSX data are from
1996-1997, and the AKARI from 2006-2007. This SED should be analysed
with caution, considering the K$_{\rm S}$-band variability mentioned
above. The outburst SED contains optical and near-infrared photometry
we obtained on 20/23 September 2010 and 16/17 November 2010. We note
that by September, the source was already $\approx$2\,mag fainter in
R-band than at maximal brightness some 20 days earlier. The shape of
the SED and the fact that the source in quiescence was practically
invisible (the only pre-outburst image where the source is visible at
optical wavelengths is the POSS2 red plate in
Fig.~\ref{fig:map_vsxj20}) suggest that the source is highly
extincted. However, dereddening its colors does not make it fall onto
the T\,Tauri locus (Fig.~\ref{fig:tcd}). Correcting for a reddening of
A$_{\rm V}$=17...20\,mag would result in a J$-$H color typical for
T\,Tauri stars, but its H$-$K$_{\rm S}$ color would still be too
red. The reason for the strange near-infrared colors of
VSX\,J205126.1+440523 may be partly interstellar reddening caused by
the small cloud in which the source is embedded and whose outlines are
visible in Fig.~\ref{fig:map_vsxj20}, partly circumstellar reddening
by an envelope or thick disk. The relative importance of these two
effects is not known, thus we do not attempt to correct for
interstellar reddening, and calculate a bolometric luminosity of
14.7\,L$_{\odot}$ by simply integrating the quiescent SED from 1.25 to
200$\,\mu$m. We calculate an outburst luminosity of 22\,L$_{\odot}$
similarly, assuming that the SED did not change above 10$\,\mu$m, and
using the September 2010 data points in Fig.~\ref{fig:sed_vsxj20}.

\section{Discussion}

\subsection{The nature of the sources in quiescence}

The SED of HBC\,722 seems to be consistent with that of a slightly
reddened T\,Tauri star, both regarding the optical--near-infrared part
of the SED and the 24$\,\mu$m photometric point. This conclusion is in
accordance with the claim of \citet{miller2010} that HBC\,722 is a
Class II object where the central star is a K7-type star. It is
noteworthy, however, that between 3.6 and 8$\,\mu$m there is an excess
emission in the SED compared to the Taurus median. In this respect,
the source somewhat resembles DR\,Tau, a highly accreting T\,Tauri
star, suggesting that HBC\,722 might be a highly accreting T\,Tauri
star even in quiescence. The fact that the quiescent optical spectrum,
taken by \citet{cohen1979} exhibits an unusually prominent H$\alpha$
emission with an equivalent width of 100\,\AA{} supports this idea.

VSXJ\,205126.1+440523 seems to be a much more reddened source. Its
quiescent bolometric luminosity ($\approx$15\,L$_{\odot}$) is
significantly larger than that of HBC\,722, indicating a somewhat
higher mass. The quiescent SED in Fig.~\ref{fig:sed_vsxj20} probably
represents a moderately reddened Class I source. The findings of
\citet{covey2010}, who determined 6\,mag$<$A$_{V}{<}$12.4\,mag from
the ratio of H emission lines observed in the outburst spectrum,
supports this idea. This scenario requires a dense envelope in the
system, but the lack of a reflection nebula around this source
suggests the lack of an extended envelope. However, high interstellar
extinction may render the scattered light invisible. The presence of
a related Herbig-Haro object also advocates for the Class I scenario.

\subsection{The outburst mechanism}

The comparison of the quiescent and the outburst SEDs
(Fig.~\ref{fig:sed_hbc722}) suggests that the brightening of HBC\,722
can be interpreted as the appearance of a hot continuum. The data
points can be fitted with a temperature of $\approx$4000\,K, somewhat
less than what one would expect from an accretion outburst, where
ionized material is present. Note however that the temperature may be
higher if the extinction towards the system is higher. It is
noteworthy that the outburst excess can be described in the
optical--near-infrared regime by a single temperature blackbody rather
than a disk-like emission reflecting a radial (usually outwards
decreasing) temperature profile. The fading of HBC\,722 is mostly
color-independent in the near-infrared regime (Fig.~\ref{fig:tcd}),
while optical colors are becoming slightly redder. This may indicate
that the hot continuum is both fading and cooling.

The reason for the brightening of VSX\,J205126.1+440523 is more
enigmatic. In this case the excess emission in outburst is not a
single-temperature radiation, but seems to have a temperature
distribution at all epochs (Fig.~\ref{fig:light_vsxj20}). The
amplitude of the outburst in the J, H and K$_{\rm S}$ bands is larger
than any brightening observed so far for YSOs. At first glance, one
may think that the brightening was due to suddenly decreased
extinction. However, in this case the source should have moved along
the reddening path in the J$-$H vs.~H$-$K$_{\rm S}$ diagram
(Fig.~\ref{fig:tcd}) which was not the case during the rising part of
the lightcurve. JHK$_{\rm S}$ photometry obtained after maximum
brightness indicate that the fading of VSX\,J205126.1+440523 initially
happened along the same path as the brightening, suggesting that
whatever was the cause of the flux changes, it was a reversible
process. Changing accretion is an appealing idea because it would
explain the presence of the Herbig-Haro object possibly ejected from
the source during a previous outburst (enhanced accretion is often
accompanied by enhanced mass outflow). It also suggests that the
outburst activity of VSXJ\,205126.1+440523 might be
repetitive. However, photometry obtained in November 2010 indicates a
deep minimum in Fig.~\ref{fig:light_vsxj20}. At the same time, the
source moved along the reddening path in Fig.~\ref{fig:tcd}. Both the
color changes and the brightness changes in the J, H, and K$_{\rm S}$
bands are consistent with an extinction increase of A$_{\rm
  V}$=9\,mag. Thus, it may be possible that the deep minimum in
November 2010 was caused by a dust condensation effect (similar to
what happened to V1515\,Cyg in 1980 \citep{kenyon1991}, or an eclipse
by dust clumps in an almost edge-on disk system (similarly to what
causes deep optical minima of the UX Orionis-type stars, see
e.g.~\citealt{eaton1995}).

\subsection{Classification as a FUor outburst}

When discovered, both sources were announced as FUor candidates.
Extensive analyses presented in \citet{semkov2010} and
\citet{miller2010} indicated that HBC\,722 can be considered as a bona
fide FUor. On the other hand, \citet{covey2010} concluded that
VSX\,J205126.1+440523 does not appear to belong to either the FUor or
the EXor class. FUors are usually bright objects in outburst with
luminosities of a few hundred L$_{\odot}$ \citep{hk96}. The immense
radiation is related to the increased accretion rate, which can reach
values up to 10$^{-4}$\,M$_{\odot}$/yr. However, the outburst
luminosities of our objects are only in the order of
10-20\,L$_{\odot}$. Assuming that the luminosity excess in eruption is
all due to the release of accretion energy, the computed accretion
rate for HBC\,722 (assuming a stellar mass of 0.5\,M$_{\odot}$ and a
radius of 3\,R$_{\odot}$) is
10$^{-6}$\,M$_{\odot}$/yr. \citet{covey2010} determined an accretion
rate of 2.5$\times$10$^{-7}$\,M$_{\odot}$/yr for
VSX\,J205126.1+440523. Both of these values are well below the typical
value for classical FUors.

The brightening and fading rates for both of our sources are also too
fast compared to classical FUor light curves. In the case of HBC\,722
we made a comparison with several young eruptive stars and found a
mismatch with the prototype FUor V1057\,Cyg but more similarities with
the light curves of EX\,Lup, the prototype of EXors, and V1647\,Ori,
an object often classified as an intermediate object between FUors and
EXors. The slow brightening of VSX\,J205126.1+440523 is not unheard of
(both FU\,Ori and V1057\,Cyg had a rise-time of about 1 year, for
other sources the estimates range between 3 and 20 years,
\citealt{bell1994}), but the fading is far too rapid. According to our
light curve (Fig.~\ref{fig:light_vsxj20}), 2.5 months after peak
brightness VSX\,J205126.1+440523 dimmed by about 5\,mag in
V-band. This is closer to the typical timescales of EXor flare-ups
than that of bona fide FUor outbursts (decades to centuries). The
non-monotonous fading of this source also resembles the light curve of
the recent outburst of EX\,Lup (Fig.~\ref{fig:context}). The moderate
resolution near-infrared spectrum obtained in outburst by
\citet{covey2010} also shows similarities to that of EX\,Lup
\citep{kospal2011}.

It is remarkable that HBC\,722 possesses all spectral characteristics
of bona fide FUors but its luminosity is an order of magnitude lower,
and its fading timescale is much faster. This suggests that the
physical mechanism which is behind the FUor-type eruptions should also
work with lower accretion rates although probably producing shorter
outbursts. This conclusion questions the thermal instability model of
\citet{bell1994}, who suggested the existence of a threshold mass
accretion rate from the outer to the inner circumstellar disk. Matter
can pile up at the inner edge of the disk and fall onto the stellar
surface following a sudden thermal instability only if the quiescent
accretion rate is higher than this threshold value. Thus, it seems
that in the regime of low luminosity outbursts
($\approx$10\,L$_{\odot}$), both FUor-like eruptions (when the source
exhibits all spectral characteristics of FUors like HBC\,722) and
EXor-like events (when the source exhibits a typical T\,Tauri spectrum
with emission lines and CO bandhead emission, somewhat similar to
VSX\,J205126.1+440523) can occur. If both HBC\,722 and
VSX\,J205126.1+440523 are indeed young eruptive stars, one might
conclude that the class of young eruptive stars is even more diverse
than what was thought before.


\begin{acknowledgements}
  This work is based in part on observations made with the Telescopio
  Carlos Sanchez operated on the island of Tenerife by the Instituto
  de Astrof\'\i{}sica de Canarias in the Observatorio del Teide. The
  authors wish to thank the telescope manager A.~Oscoz, support
  astronomer P.~Monta\~nes, and telescope operators R.~Mart\'\i{}, and
  M.~D\'\i{}az for their help during the observations. This work is
  based in part on observations made with the Spitzer Space Telescope,
  which is operated by the Jet Propulsion Laboratory, California
  Institute of Technology under a contract with NASA. This work is
  based in part on data obtained as part of the UKIRT Infrared Deep
  Sky Survey. This publication makes use of data products from the Two
  Micron All Sky Survey, which is a joint project of the University of
  Massachusetts and the Infrared Processing and Analysis
  Center/California Institute of Technology, funded by the NASA and
  the National Science Foundation. The research of \'A.K. is supported
  by the Nederlands Organization for Scientific Research.
\end{acknowledgements}


\bibliographystyle{aa}
\bibliography{paper}{}

\begin{thebibliography}{27}
\expandafter\ifx\csname natexlab\endcsname\relax\def\natexlab#1{#1}\fi

\bibitem[{{{\'A}brah{\'a}m} {et~al.}(2009){{\'A}brah{\'a}m}, {Juh{\'a}sz},
  {Dullemond}, {K{\'o}sp{\'a}l}, {van Boekel}, {Bouwman}, {Henning},
  {Mo{\'o}r}, {Mosoni}, {Sicilia-Aguilar}, \& {Sipos}}]{abraham2009}
{{\'A}brah{\'a}m}, P., {Juh{\'a}sz}, A., {Dullemond}, C.~P., {et~al.} 2009,
  \nat, 459, 224

\bibitem[{{Acosta-Pulido} {et~al.}(2007){Acosta-Pulido}, {Kun},
  {{\'A}brah{\'a}m}, {K{\'o}sp{\'a}l}, {Csizmadia}, {Kiss}, {Mo{\'o}r},
  {Szabados}, {Benk{\H o}}, {Barrena Delgado}, {Charcos-Llorens}, {Eredics},
  {Kiss}, {Manchado}, {R{\'a}cz}, {Ramos Almeida}, {Sz{\'e}kely}, \&
  {Vidal-N{\'u}{\~n}ez}}]{acosta2007}
{Acosta-Pulido}, J.~A., {Kun}, M., {{\'A}brah{\'a}m}, P., {et~al.} 2007, \aj,
  133, 2020

\bibitem[{{Bally} \& {Reipurth}(2003)}]{bally2003}
{Bally}, J. \& {Reipurth}, B. 2003, \aj, 126, 893

\bibitem[{{Bell} \& {Lin}(1994)}]{bell1994}
{Bell}, K.~R. \& {Lin}, D.~N.~C. 1994, \apj, 427, 987

\bibitem[{{Cardelli} {et~al.}(1989){Cardelli}, {Clayton}, \&
  {Mathis}}]{cardelli1989}
{Cardelli}, J.~A., {Clayton}, G.~C., \& {Mathis}, J.~S. 1989, \apj, 345, 245

\bibitem[{{Cohen} \& {Kuhi}(1979)}]{cohen1979}
{Cohen}, M. \& {Kuhi}, L.~V. 1979, \apjs, 41, 743

\bibitem[{{Covey} {et~al.}(2011){Covey}, {Hillenbrand}, {Miller}, {Poznanski},
  {Cenko}, {Silverman}, {Bloom}, {Kasliwal}, {Fischer}, {Rayner}, {Rebull},
  {Butler}, {Filippenko}, {Law}, {Ofek}, {Ag{\"u}eros}, {Dekany}, {Rahmer},
  {Hale}, {Smith}, {Quimby}, {Nugent}, {Jacobsen}, {Zolkower}, {Velur},
  {Walters}, {Henning}, {Bui}, {McKenna}, {Kulkarni}, \& {Klein}}]{covey2010}
{Covey}, K.~R., {Hillenbrand}, L.~A., {Miller}, A.~A., {et~al.} 2011, \aj, 141,
  40

\bibitem[{{Cutri} {et~al.}(2003){Cutri}, {Skrutskie}, {van Dyk}, {Beichman},
  {Carpenter}, {Chester}, {Cambresy}, {Evans}, {Fowler}, {Gizis}, {Howard},
  {Huchra}, {Jarrett}, {Kopan}, {Kirkpatrick}, {Light}, {Marsh}, {McCallon},
  {Schneider}, {Stiening}, {Sykes}, {Weinberg}, {Wheaton}, {Wheelock}, \&
  {Zacarias}}]{cutri2003}
{Cutri}, R.~M., {Skrutskie}, M.~F., {van Dyk}, S., {et~al.} 2003, {2MASS All
  Sky Catalog of point sources.}

\bibitem[{{D'Alessio} {et~al.}(1999){D'Alessio}, {Calvet}, {Hartmann},
  {Lizano}, \& {Cant{\'o}}}]{dalessio1999}
{D'Alessio}, P., {Calvet}, N., {Hartmann}, L., {Lizano}, S., \& {Cant{\'o}}, J.
  1999, \apj, 527, 893

\bibitem[{{Eaton} \& {Herbst}(1995)}]{eaton1995}
{Eaton}, N.~L. \& {Herbst}, W. 1995, \aj, 110, 2369

\bibitem[{{Egan} {et~al.}(2003){Egan}, {Price}, {Kraemer}, {Mizuno}, {Carey},
  {Wright}, {Engelke}, {Cohen}, \& {Gugliotti}}]{egan2003}
{Egan}, M.~P., {Price}, S.~D., {Kraemer}, K.~E., {et~al.} 2003, VizieR Online
  Data Catalog, 5114, 0

\bibitem[{{Furlan} {et~al.}(2006){Furlan}, {Hartmann}, {Calvet}, {D'Alessio},
  {Franco-Hern{\'a}ndez}, {Forrest}, {Watson}, {Uchida}, {Sargent}, {Green},
  {Keller}, \& {Herter}}]{furlan2006}
{Furlan}, E., {Hartmann}, L., {Calvet}, N., {et~al.} 2006, \apjs, 165, 568

\bibitem[{{Gieseking}(1973)}]{gieseking1973}
{Gieseking}, F. 1973, Information Bulletin on Variable Stars, 806, 1

\bibitem[{{Hartmann} \& {Kenyon}(1996)}]{hk96}
{Hartmann}, L. \& {Kenyon}, S.~J. 1996, \araa, 34, 207

\bibitem[{{Ishihara} {et~al.}(2010){Ishihara}, {Onaka}, {Kataza}, {Salama},
  {Alfageme}, {Cassatella}, {Cox}, {Garc{\'{\i}}a-Lario}, {Stephenson},
  {Cohen}, {Fujishiro}, {Fujiwara}, {Hasegawa}, {Ita}, {Kim}, {Matsuhara},
  {Murakami}, {M{\"u}ller}, {Nakagawa}, {Ohyama}, {Oyabu}, {Pyo}, {Sakon},
  {Shibai}, {Takita}, {Tanab{\'e}}, {Uemizu}, {Ueno}, {Usui}, {Wada},
  {Watarai}, {Yamamura}, \& {Yamauchi}}]{ishihara2010}
{Ishihara}, D., {Onaka}, T., {Kataza}, H., {et~al.} 2010, \aap, 514, A1

\bibitem[{{Itagaki} \& {Yamaoka}(2010)}]{itagaki2010}
{Itagaki}, K. \& {Yamaoka}, H. 2010, Central Bureau Electronic Telegrams, 2426,
  1

\bibitem[{{Juh{\'a}sz} {et~al.}(2011){Juh{\'a}sz}, {Dullemond}, {van Boekel},
  {Bouwman}, {{\'A}brah{\'a}m}, {Acosta-Pulido}, {Henning}, {K{\'o}sp{\'a}l},
  {Sicilia-Aguilar}, {Jones}, {Mo{\'o}r}, {Mosoni}, {Reg{\'a}ly}, {Szokoly}, \&
  Sipos}]{juhasz2010}
{Juh{\'a}sz}, A., {Dullemond}, C.~P., {van Boekel}, R., {et~al.} 2011, \apj,
  submitted

\bibitem[{{Kenyon} {et~al.}(1991){Kenyon}, {Hartmann}, \&
  {Kolotilov}}]{kenyon1991}
{Kenyon}, S.~J., {Hartmann}, L.~W., \& {Kolotilov}, E.~A. 1991, \pasp, 103,
  1069

\bibitem[{{Koornneef}(1983)}]{koornneef1983}
{Koornneef}, J. 1983, \aap, 128, 84

\bibitem[{{K{\'o}sp{\'a}l} {et~al.}(in prep.){K{\'o}sp{\'a}l},
  {{\'A}brah{\'a}m}, {Reg{\'a}ly}, {Henning}, {Goto}, \&
  {Juh{\'a}sz}}]{kospal2011}
{K{\'o}sp{\'a}l}, {\'A}., {{\'A}brah{\'a}m}, P., {Reg{\'a}ly}, Z., {et~al.} in
  prep., \apj

\bibitem[{{K{\'o}sp{\'a}l} {et~al.}(2010){K{\'o}sp{\'a}l}, {Salter},
  {Hogerheijde}, {Mo{\'o}r}, \& {Blake}}]{kospal2010}
{K{\'o}sp{\'a}l}, {\'A}., {Salter}, D.~M., {Hogerheijde}, M.~R., {Mo{\'o}r},
  A., \& {Blake}, G.~A. 2010, ArXiv e-prints

\bibitem[{{Meyer} {et~al.}(1997){Meyer}, {Calvet}, \&
  {Hillenbrand}}]{meyer1997}
{Meyer}, M.~R., {Calvet}, N., \& {Hillenbrand}, L.~A. 1997, \aj, 114, 288

\bibitem[{{Miller} {et~al.}(2010){Miller}, {Hillenbrand}, {Covey}, {Poznanski},
  {Silverman}, {Kleiser}, {Royas-Ayala}, {Muirhead}, {Cenko}, {Bloom},
  {Kasliwal}, {Filippenko}, {Law}, {Ofek}, {Dekany}, {Rahmer}, {Hale}, {Smith},
  {Quimby}, {Nugent}, {Jacobsen}, {Zolkower}, {Velur}, {Walters}, {Henning},
  {Bui}, {McKenna}, {Kulkarni}, \& {Klein}}]{miller2010}
{Miller}, A.~A., {Hillenbrand}, L.~A., {Covey}, K.~R., {et~al.} 2010, \apj,
  submitted

\bibitem[{{Munari} {et~al.}(2010){Munari}, {Valisa}, {Dallaporta}, \&
  {Itagaki}}]{munari2010}
{Munari}, U., {Valisa}, P., {Dallaporta}, S., \& {Itagaki}, K. 2010, Central
  Bureau Electronic Telegrams, 2428, 1

\bibitem[{{Semkov} \& {Peneva}(2010)}]{semkov2010a}
{Semkov}, E. \& {Peneva}, S. 2010, The Astronomer's Telegram, 2801, 1

\bibitem[{{Semkov} {et~al.}(2010){Semkov}, {Peneva}, {Munari}, {Milani}, \&
  {Valisa}}]{semkov2010}
{Semkov}, E.~H., {Peneva}, S.~P., {Munari}, U., {Milani}, A., \& {Valisa}, P.
  2010, \aap, 523, L3

\bibitem[{{Straizys} {et~al.}(1989){Straizys}, {Meistas}, {Vansevicius}, \&
  {Goldberg}}]{straizys1989}
{Straizys}, V., {Meistas}, E., {Vansevicius}, V., \& {Goldberg}, E.~P. 1989,
  \aap, 222, 82

\end{thebibliography}


\Online
\vspace*{70mm}
\hspace*{65mm}
\mbox{\LARGE{Online Material}}

\onlfig{1}{
\begin{figure*}
\includegraphics[width=\columnwidth]{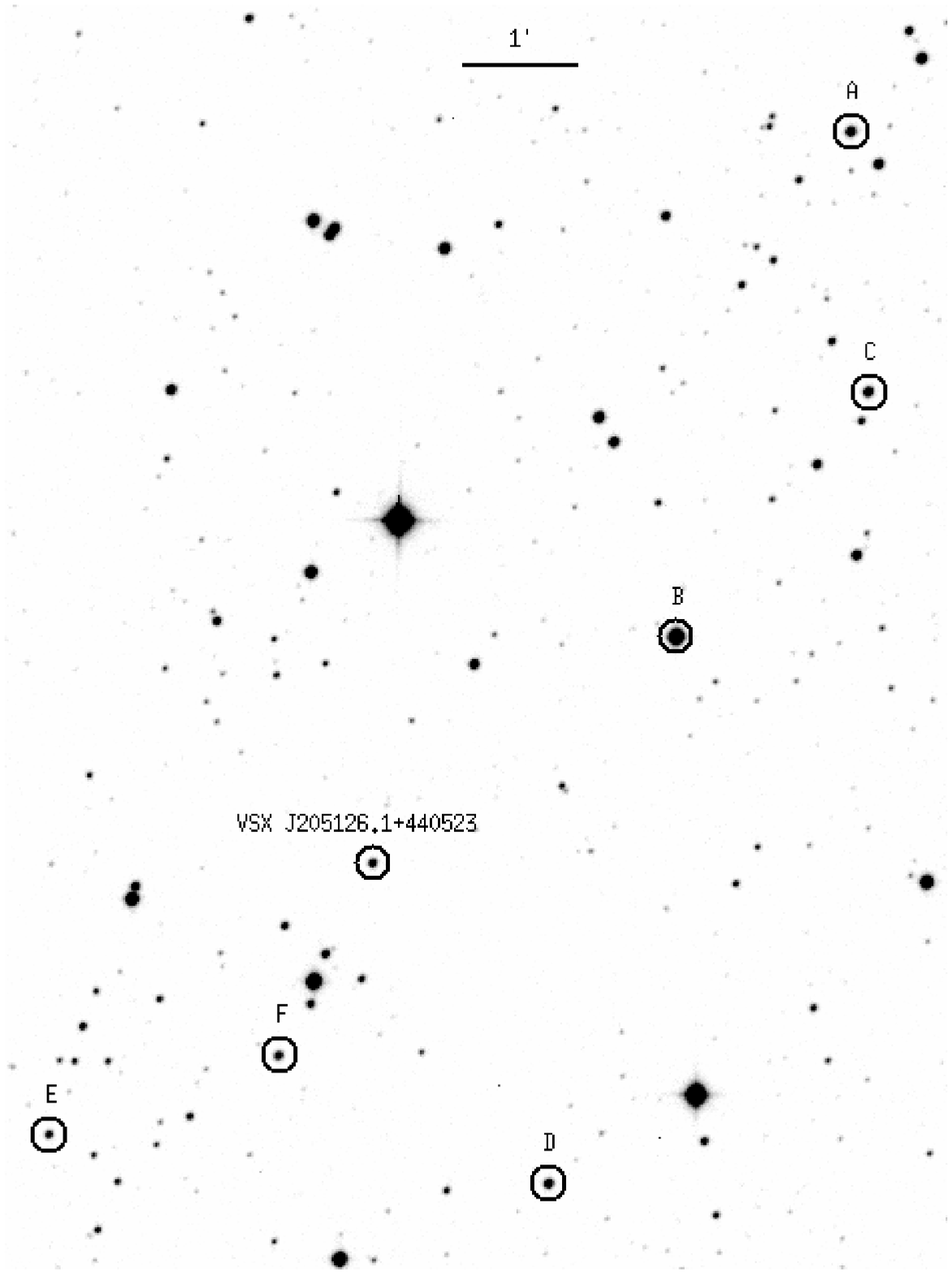}
\caption{Finding chart for VSX\,J205126.1+440523 and its comparison
  stars. North is up and east is to the left.}\label{fig:compstars}
\end{figure*}
}

\onltab{1}{
\begin{table*}
\caption{V(RI)$_{\rm C}$ photometry for the comparison stars used
  for VSX\,J205126.1+440523. Typical uncertainties are
  0.04\,mag.}\label{tab:compstars}
\begin{tabular}{ccccc}
\hline \hline
Star & 2MASS id & V & R$_{\rm C}$ & I$_{\rm C}$ \\
\hline
A & 2MASS J20510393+4411406 & 14.091 & 13.509 & 12.985 \\
B & 2MASS J20511195+4407213 & 14.268 & 12.400 & 10.968 \\
C & 2MASS J20510294+4409277 & 14.684 & 13.856 & 13.198 \\
D & 2MASS J20511769+4402412 & 14.981 & 14.146 & 13.307 \\
E & 2MASS J20514142+4403032 & 15.300 & 14.790 & 14.236 \\
F & 2MASS J20513057+4403449 & 15.924 & 14.864 & 13.816 \\
\hline
\end{tabular}
\end{table*}
}

\onltab{2}{
\begin{table*}
\caption{Photometry for VSX\,J205126.1+440523.}\label{tab:phot_vsxj20}
\begin{tabular}{ccccccccc}
\hline \hline
Date & JD$\,{-}\,$2,400,000 & V & R & I & J & H & K$_{\rm S}$ & Telescope\\
\hline
2006-Jul-20 & 53936.95 &           &           &           & 19.34(7) & 16.50(7) & 13.15(11)& UKIRT   \\
\hline
2010-Sep-19 & 55459.47 &           &           &           & 11.41(1) & 9.85(2)  & 8.35(4)  & TCS     \\
2010-Sep-20 & 55460.31 & 16.76(5)  & 15.19(4)  & 13.71(4)  &          &          &          & Schmidt \\
2010-Sep-20 & 55460.46 &           &           &           & 11.46(1) & 9.80(1)  & 8.25(2)  & TCS     \\
2010-Sep-21 & 55461.32 & 16.74(4)  & 15.16(4)  & 13.69(3)  &          &          &          & Schmidt \\
2010-Sep-23 & 55463.36 & 16.79(5)  & 15.30(4)  & 13.82(4)  &          &          &          & Schmidt \\
2010-Sep-24 & 55464.27 &           & 15.12(8)  & 13.75(4)  &          &          &          & Schmidt \\
2010-Oct-09 & 55478.47 & 16.35(4)  & 14.80(4)  & 13.41(4)  &          &          &          & Schmidt \\
2010-Oct-10 & 55479.34 & 16.39(4)  & 14.83(4)  & 13.40(4)  &          &          &          & Schmidt \\
2010-Oct-11 & 55480.34 & 16.33(4)  & 14.69(4)  & 13.38(4)  &          &          &          & Schmidt \\
2010-Oct-12 & 55481.47 & 16.56(4)  & 14.94(4)  & 13.56(4)  &          &          &          & Schmidt \\
2010-Oct-13 & 55482.31 & 16.67(4)  & 15.08(5)  & 13.67(4)  &          &          &          & Schmidt \\
2010-Oct-14 & 55483.30 & 16.67(5)  & 14.99(4)  & 13.86(6)  &          &          &          & Schmidt \\
2010-Nov-04 & 55505.32 &           &           &           & 12.75(3) & 10.82(2) & 8.90(1)  & TCS     \\
2010-Nov-04 & 55505.32 & 18.15(5)  & 16.51(6)  & 15.08(5)  &          &          &          & IAC-80  \\
2010-Nov-05 & 55506.35 &           &           &           & 12.60(3) & 10.78(2) & 8.98(3)  & TCS     \\
2010-Nov-05 & 55506.35 & 17.98(5)  & 16.40(4)  & 14.84(4)  &          &          &          & IAC-80  \\
2010-Nov-06 & 55507.35 &           &           &           & 13.13(1) & 11.08(1) & 9.18(3)  & TCS     \\
2010-Nov-06 & 55507.35 &           & 17.14(10) & 15.43(11) &          &          &          & IAC-80  \\
2010-Nov-07 & 55508.34 &           &           &           & 13.04(1) & 11.05(1) & 9.20(3)  & TCS     \\
2010-Nov-07 & 55508.40 & 18.87(11) & 17.21(10) & 15.56(9)  &          &          &          & IAC-80  \\
2010-Nov-13 & 55514.40 & 19.55(13) & 18.01(17) & 16.47(11) &          &          &          & Schmidt \\
2010-Nov-14 & 55515.37 & 20.46(30) & 18.46(7)  & 16.76(4)  &          &          &          & Schmidt \\
2010-Nov-15 & 55516.20 & 20.05(25) & 18.72(7)  & 16.84(5)  &          &          &          & Schmidt \\
2010-Nov-16 & 55517.42 &           &           &           &          &          &          & TCS     \\
2010-Nov-17 & 55518.42 &           &           &           &          &          &          & TCS     \\
2010-Nov-18 & 55519.42 &           &           &           &          &          &          & TCS     \\
2010-Nov-18 & 55524.20 &           &           & 15.97(4)  &          &          &          & Schmidt \\
2010-Nov-30 & 55531.24 & 19.71(14) & 17.93(8)  & 16.05(4)  &          &          &          & Schmidt \\
2010-Dec-10 & 55541.27 & 17.79(5)  & 16.06(5)  & 14.43(4)  &          &          &          & Schmidt \\
\hline
\end{tabular}
\end{table*}
}

\onltab{3}{
\begin{table*}
\caption{Photometry for HBC\,722.}\label{tab:phot_hbc722}
\begin{tabular}{cccccccccc}
\hline \hline
Date & JD$\,{-}\,$2,400,000 &  B   &     V     &    R      &    I      &   J       &   H      & K$_{\rm S}$ & Telescope\\
\hline
2006-Jul-20 & 53936.99 &           &           &           &           & 13.23(8) & 12.21(6) & 11.32(7) & UKIRT   \\
\hline
2010-Sep-19 & 55459.44 &           &           &           &           & 9.96(5)  & 9.12(1)  & 8.68(3)  & TCS     \\
2010-Sep-20 & 55460.28 &           & 13.71(3)  & 12.68(1)  & 11.67(1)  &          &          &          & Schmidt \\
2010-Sep-20 & 55460.44 &           &           &           &           & 9.99(1)  & 9.09(1)  & 8.70(2)  & TCS     \\
2010-Sep-21 & 55461.29 &           & 13.76(2)  & 12.71(2)  & 11.70(1)  &          &          &          & Schmidt \\
2010-Sep-23 & 55463.37 &           & 13.70(11) & 12.59(2)  & 11.60(1)  &          &          &          & Schmidt \\
2010-Sep-24 & 55464.32 &           & 13.62(9)  & 12.59(5)  & 11.65(4)  &          &          &          & Schmidt \\
2010-Oct-09 & 55478.43 & 15.23(2)  & 13.70(2)  & 12.68(2)  & 11.67(1)  &          &          &          & Schmidt \\
2010-Oct-14 & 55483.44 &           & 13.85(1)  & 12.80(1)  & 11.76(1)  &          &          &          & Schmidt \\
2010-Nov-04 & 55505.35 &           &           &           &           & 10.16(3) & 9.28(1)  & 8.75(1)  & TCS     \\
2010-Nov-04 & 55505.35 & 15.46(5)  & 13.96(3)  & 12.89(2)  & 11.86(1)  &          &          &          & IAC-80  \\
2010-Nov-05 & 55506.37 &           &           &           &           & 10.18(3) & 9.22(3)  & 8.75(6)  & TCS     \\
2010-Nov-05 & 55506.39 & 15.47(2)  & 13.96(1)  & 12.91(2)  & 11.85(6)  &          &          &          & IAC-80  \\
2010-Nov-06 & 55507.35 & 15.50(5)  & 13.97(5)  & 12.92(6)  &           &          &          &          & IAC-80  \\
2010-Nov-06 & 55507.37 &           &           &           &           & 10.16(1) & 9.26(1)  & 8.71(1)  & TCS     \\
2010-Nov-07 & 55508.36 &           &           &           &           & 10.21(1) & 9.27(1)  & 8.78(1)  & TCS     \\
2010-Nov-07 & 55508.38 & 15.56(5)  & 14.02(2)  & 13.01(3)  &           &          &          &          & IAC-80  \\
2010-Nov-13 & 55514.38 & 15.57(5)  & 14.07(1)  & 13.00(1)  & 11.97(3)  &          &          &          & Schmidt \\
2010-Nov-14 & 55515.41 & 15.68(4)  & 14.07(2)  & 13.02(1)  & 11.98(1)  &          &          &          & Schmidt \\
2010-Nov-15 & 55516.25 & 15.66(3)  & 14.07(1)  & 13.05(2)  & 12.00(1)  &          &          &          & Schmidt \\
2010-Nov-16 & 55517.41 & 15.69(3)  & 14.13(2)  & 13.10(1)  &           &          &          &          & IAC-80  \\
2010-Nov-16 & 55517.42 &           &           &           &           & 10.28(1) & 9.38(1)  & 8.81(3)  & TCS     \\
2010-Nov-17 & 55518.42 &           &           &           &           & 10.31(1) & 9.37(1)  & 8.86(1)  & TCS     \\
2010-Nov-18 & 55519.44 &           &           &           &           & 10.31(1) & 9.36(1)  & 8.82(1)  & TCS     \\
2010-Nov-18 & 55519.44 & 15.73(5)  & 14.16(5)  & 13.16(5)  &           &          &          &          & IAC-80  \\
2010-Nov-19 & 55520.40 &           &           &           &           & 10.24(1) & 9.34(1)  & 8.81(1)  & TCS     \\
2010-Nov-19 & 55520.40 & 15.69(9)  & 14.15(3)  & 13.10(3)  &           &          &          &          & IAC-80  \\
2010-Nov-23 & 55524.33 &           &           &           & 12.09(5)  &          &          &          & RCC     \\
2010-Nov-30 & 55531.19 & 15.79(5)  & 14.23(2)  & 13.18(1)  & 12.12(1)  &          &          &          & Schmidt \\
2010-Dec-01 & 55532.31 & 15.79(4)  & 14.26(2)  & 13.20(1)  & 12.14(1)  &          &          &          & IAC-80  \\
2010-Dec-03 & 55534.42 & 15.80(3)  & 14.26(2)  & 13.22(1)  & 12.14(2)  &          &          &          & IAC-80  \\
2010-Dec-10 & 55541.31 & 15.93(4)  & 14.39(2)  & 13.30(2)  & 12.23(2)  &          &          &          & Schmidt \\
2010-Dec-12 & 55543.34 & 15.98(5)  & 14.40(2)  & 13.33(2)  & 12.23(2)  &          &          &          & IAC-80 \\
2011-Jan-02 & 55564.32 & 16.17(4)  & 14.59(2)  & 13.52(2)  & 12.40(2)  &          &          &          & IAC-80 \\
\hline
\end{tabular}
\end{table*}
}

\onlfig{2}{
\begin{figure*}
\centering
\includegraphics[height=1.75\columnwidth,angle=90]{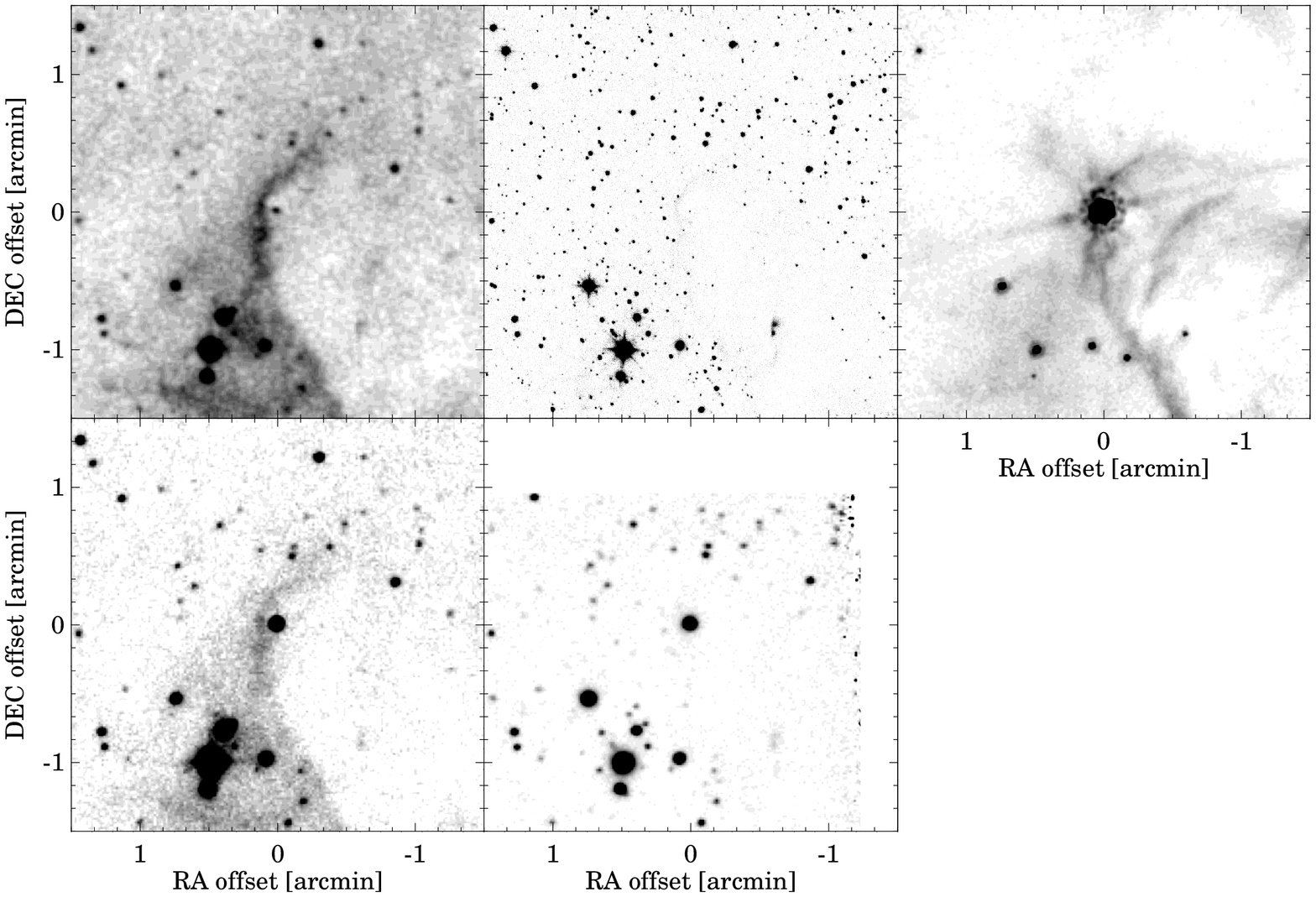}
\caption{VSX\,J205126.1+440523 and its surroundings. The field of view
  is 3$'\times$3$'$. The upper row shows pre-outburst images (left:
  POSS2 red, middle: UKIDSS J, right: Spitzer/IRAC 8$\,\mu$m), while
  the bottom row shows outburst images (left: Konkoly Schmidt, middle:
  Teide TCS).}
\label{fig:map_vsxj20}
\end{figure*}
}

\onlfig{3}{
\begin{figure*}
\centering
\includegraphics[height=1.75\columnwidth,angle=90]{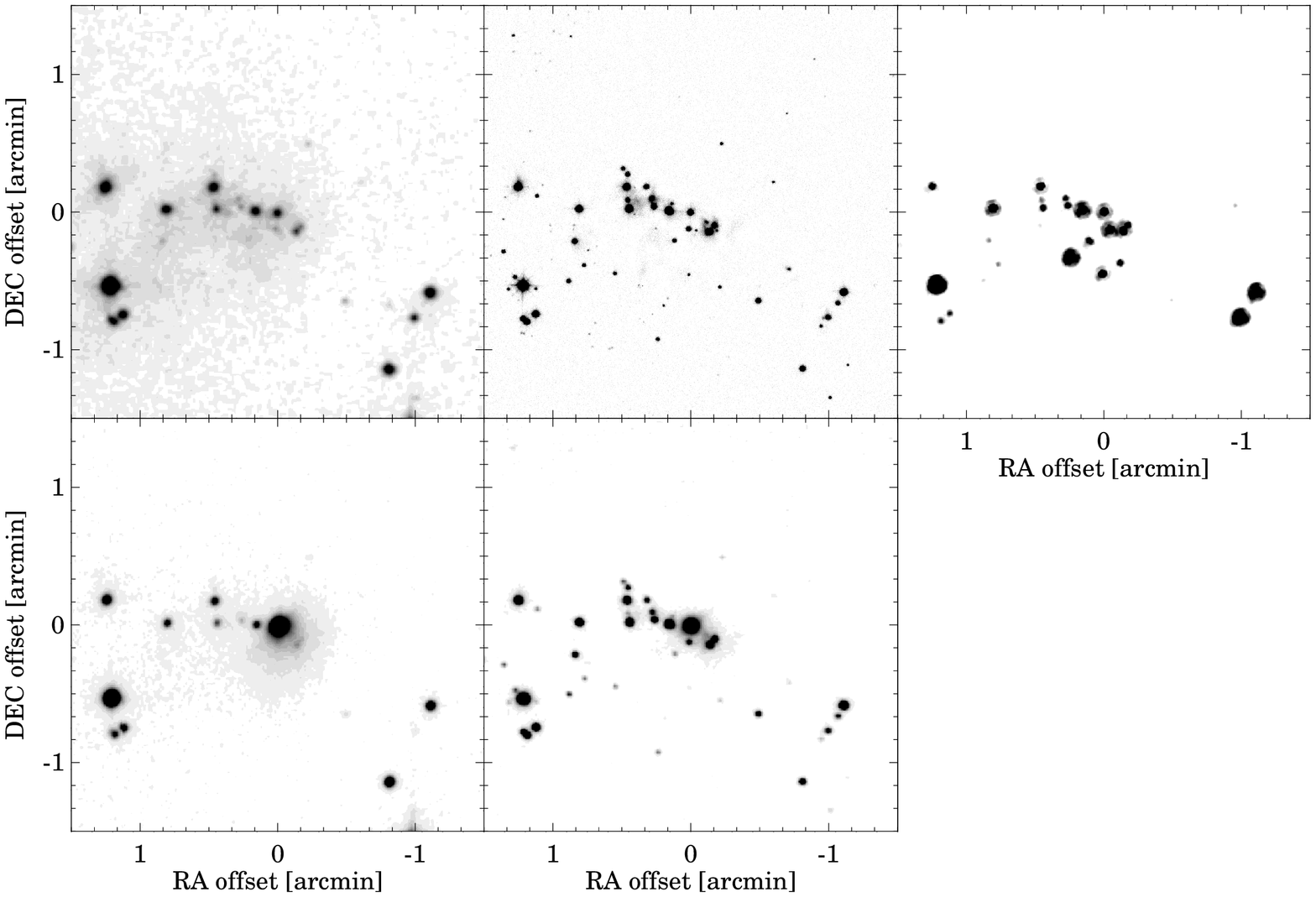}
\caption{HBC\,722 and its surroundings. The field of view is
  3$'\times$3$'$. The uppper row shows pre-outburst images (left:
  POSS2 red, middle: UKIDSS J, right: Spitzer/IRAC 8$\,\mu$m), while
  the bottom row shows outburst images (left: Konkoly Schmidt, middle:
  Teide TCS).}
\label{fig:map_hbc722}
\end{figure*}
}

\end{document}